\documentclass[twocolumn,a4paper]{aastex7}
\usepackage{lineno}
\usepackage{multirow}
\usepackage{booktabs}

\usepackage{amsmath,latexsym}    % need for subequations
\usepackage{graphicx}   % need for figures
\usepackage{verbatim}   % useful for program listings
\usepackage{color}      % use if color is used in text
\usepackage{subfigure}  % use for side-by-side figures
\usepackage{hyperref}   % use for hypertext links, including those to external documents and URLs
\usepackage{tensor}     %use for tensor, as it should be.
\usepackage{amssymb}
\usepackage{enumerate}	% can be used to enumerate several items
\usepackage{braket}		% for braket notation

\received{}
\revised{}
\accepted{}
\published{}

\shortauthors{Roy Choudhury}

\begin{document}

\title{Updated 1.1\% Precision Values of the Hubble Constant with Corrected Pantheon+ and Dark Energy Survey (DES)-DOVEKIE Type Ia Supernovae}

\author[0000-0002-5589-3454]{Shouvik Roy Choudhury}
\affiliation{Institute of Astronomy and Astrophysics, Academia Sinica, No. 1, Section 4, Roosevelt Road, Taipei 106319, Taiwan}
\email[show]{sroy@asiaa.sinica.edu.tw}

\begin{abstract}
We update the Hubble constant using corrected Pantheon+ and DES-DOVEKIE Type Ia supernovae, presenting the first application of DES-DOVEKIE to a local distance-ladder determination of $H_0$. Recent reanalyses have brought Pantheon+, DES-DOVEKIE, and Union3.1 into substantially closer agreement on dynamical dark energy, with $3.2$--$3.4\sigma$ preferences when combined with DESI BAO and CMB measurements. This makes a corresponding reassessment of the local distance scale timely. The $H_0$ Distance Network (H0DN) baseline is $73.499\pm0.809\,\mathrm{km\,s^{-1}\,Mpc^{-1}}$, in $7.1\sigma$ tension with the CMB+$\Lambda$CDM value, $67.24\pm0.35\,\mathrm{km\,s^{-1}\,Mpc^{-1}}$. A revised Pantheon+ host-mass correction changes 114 of 1701 apparent-magnitude rows, including 52 of 277 Hubble-flow rows but no calibrator supernovae, leaving the calibrator rung unchanged. Applying only these magnitude corrections, while retaining the H0DN covariance, redshifts, and velocities, gives our primary result, $H_0=73.264\pm0.806\,\mathrm{km\,s^{-1}\,Mpc^{-1}}$, corresponding to a $6.85\sigma$ tension. Replacing the complete Hubble-flow treatment with corrected Pantheon+ instead gives $73.139\pm0.791\,\mathrm{km\,s^{-1}\,Mpc^{-1}}$ and $6.82\sigma$. The first calculation isolates the revised magnitudes, whereas the second consistently replaces the complete supernova rung, including its covariance, redshifts, and velocity treatment. For DES-DOVEKIE, 220 supernovae span $0.025\le z_{\rm HD}\le0.151$, and 196 overlap corrected Pantheon+. These overlaps determine the offset that places the public DES-DOVEKIE distance moduli on the H0DN apparent-magnitude scale. Holding the offset fixed gives $H_0=73.267\pm0.828\,\mathrm{km\,s^{-1}\,Mpc^{-1}}$ and $6.70\sigma$; propagating its available uncertainty and covariance gives $73.175\pm0.854\,\mathrm{km\,s^{-1}\,Mpc^{-1}}$ and $6.43\sigma$. Extending either Hubble-flow sample to $z_{\rm HD}\simeq0.3$ raises rather than lowers $H_0$, leaving the CMB comparison essentially unchanged. Thus, supernova revisions important for current dynamical-dark-energy constraints do not resolve the Hubble tension.
\end{abstract}

\section{Introduction}

The Hubble constant, $H_0$, is the present expansion rate of the Universe.  Measurements made through the local distance ladder remain higher than values inferred from cosmic microwave background observations within the standard cosmological model \citep{Verde:2019ivm,DiValentino:2021izs,Kamionkowski:2022pkx, CosmoVerseNetwork:2025alb, Vagnozzi:2023nrq, Cai:2026swf}.  The Supernovae, $H_0$, for the Equation of State of dark energy (SH0ES) distance ladder gives a value near 73 km s$^{-1}$ Mpc$^{-1}$ \citep{Riess:2021jrx}, while independent late-Universe routes use the tip of the red-giant branch, megamasers, or surface-brightness fluctuations \citep{Freedman:2024eph,Pesce:2020xfe,Blakeslee:2021rqi}.  The H0 Distance Network (H0DN) combines these routes and their shared calibrations in one public fit.  Its baseline value is $73.499\pm0.809$ km s$^{-1}$ Mpc$^{-1}$ \citep{H0DN:2025lyy}.

Type Ia supernovae connect nearby galaxies with geometric distance measurements to galaxies far enough away that cosmic expansion dominates their individual motions.  Their luminosities are standardized through the observed light-curve width and color, with additional dependence on the host galaxy \citep{Phillips:1993ng,Tripp:1997wt,SNLS:2007cqk,Brout:2020msh}.  A change in the standardized magnitudes, redshifts, or covariance of this Hubble-flow rung can therefore change $H_0$, even when the nearby calibrators are left untouched.

Two recent data products motivate a direct update of this rung.  First, a self-consistent remeasurement of supernova host galaxies changes the host-property correction for part of Pantheon+ \citep{Hoyt:2026fve}, whose public data contain 1701 light curves \citep{Brout:2022vxf,Scolnic:2021amr}.  We refer to the resulting file simply as corrected Pantheon+.  Second, the Dark Energy Survey five-year supernova sample has been recalibrated as DES-DOVEKIE, with a new Hubble diagram and covariance \citep{DES:2024jxu,DES:2024hip,DES:2025sig}.  The two data products do not enter H0DN in the same form: corrected Pantheon+ supplies standardized apparent magnitudes, whereas DES-DOVEKIE supplies distance moduli. To our knowledge, this is the first use of DES-Dovekie in a
local distance-ladder determination of $H_0$.

These revised supernova products are especially timely because the corrected Pantheon+, Union3.1, and DES-DOVEKIE samples now give substantially more consistent constraints on dynamical dark energy (evidence of dynamical dark energy ranging from 3.2-3.4$\sigma$) when combined with DESI BAO and CMB measurements \citep{Hoyt:2026fve}. Because the same low-redshift supernova measurements also define the Hubble-flow rung of local distance-ladder analyses, it is important to determine whether the corrections that improve cross-sample dark-energy consistency materially alter the inferred Hubble constant. An analogous determination cannot yet be made with Union3.1 because, as of this writing, its public data products do not provide the individual-supernova apparent magnitudes, redshifts, and peculiar-velocity corrections required to construct the Hubble-flow rung. 

The structure of the paper is as follows: Section~\ref{sec:2} outlines the analysis methodology. In Section~\ref{sec:3}, we present and discuss the results of our statistical analysis. We conclude in Section~\ref{sec:4}.

\section{Data and Method} \label{sec:2}

\subsection{The H0 Distance Network}

We use the public H0DN calculation files throughout this analysis.\footnote{\url{https://github.com/StefCas789/H0DN}}

The public H0DN baseline contains 55 calibrator supernovae and 277 Hubble-flow rows.  Some supernova names occur more than once because separate standardized measurements are retained as separate rows.  H0DN first uses the Hubble-flow sample to determine an intercept, denoted by $\alpha_B$, and combines it with the calibrated SN Ia absolute magnitude, $M_B$, through
\begin{equation}
  \log_{10} H_0 = 0.2 M_B + \alpha_B + 5.
  \label{eq:h0dn}
\end{equation}
The complete network also contains distance indicators that do not use the supernova Hubble-flow rung \citep{H0DN:2025lyy}.  For this reason, the magnitude change needed to move the full network is not given by the supernova intercept alone.  An exact H0DN rerun shows that a coherent faintening of all baseline Hubble-flow magnitudes by $0.236$ mag is required to move the network central value from $73.499$ to $67.24$ km s$^{-1}$ Mpc$^{-1}$.  The supernova-only conversion, $5\log_{10}(73.499/67.24)=0.193$ mag, is smaller because it omits the other distance indicators.

All runs in this paper use $q_0=-0.55$ and $j_0=1$, matching the H0DN baseline.  Here $q_0$ is the present deceleration parameter, which describes the acceleration of the expansion, and $j_0$ is the present jerk parameter, which describes its next time derivative.  Fixing both values makes every comparison below a change of supernova data rather than a change of the assumed expansion history.

\subsection{Corrected Pantheon+}

We first isolate the revised Pantheon+ magnitude correction. We independently reconstructed the corrected Pantheon+ data file from public inputs and verified that its numerical contents agree, to floating-point precision, with the file privately provided to us by \citet{Hoyt:2026fve}. We reconstruct the corrected magnitude vector from the public Pantheon+ table using the prescription in Appendix~F of \citet{Hoyt:2026fve}.  The reconstructed standardized magnitudes for the 114 affected rows of the corrected Pantheon+ table and the scripts used for their reconstruction are archived on Zenodo under a Creative Commons Attribution 4.0 license \citep{RoyChoudhury2026Zenodo}:\dataset[doi:10.5281/zenodo.22017350]{https://doi.org/10.5281/zenodo.22017350}. \footnote{The public Pantheon+ input table, the independently generated corrected table, the complete reconstruction script, a row-by-row audit are also available on github: \url{https://github.com/shouvikrc/pantheonplus-hostmass-correction-reconstruction}}.

For the 713 public Pantheon+ rows satisfying $0.01\leq z_{\rm CMB}<0.15$, we represent the Pantheon+ magnitude bias correction as a function of the SALT color parameter $c_{\rm SALT}$ separately below and above the host-mass step at $\log_{10}(M_*/M_\odot)=10$.  Here $z_{\rm CMB}$ is the redshift in the cosmic-microwave-background frame, $M_*$ is the host-galaxy stellar mass, and $M_\odot$ is the solar mass.  The two relations are denoted by $f_{\rm low}(c_{\rm SALT})$ and $f_{\rm high}(c_{\rm SALT})$.  Separate unweighted degree-three polynomial fits to the public \texttt{biasCor\_m\_b} values numerically reproduce the low- and high-mass spline representations described by \citet{Hoyt:2026fve}.  Out of the 713 rows, 114 rows required this update because their published low-redshift host masses were systematically underestimated by approximately $0.6$ dex, causing the revised masses to move these supernovae across the host-mass division at $\log_{10}(M_*/M_\odot)=10$ used in the Pantheon+ bias correction. For these 114 rows with the published host mass in $9.4\leq\log_{10}(M_*/M_\odot)<10$, which cross the mass step after the 0.6 dex host-mass revision, we calculate
\begin{equation}
	\Delta_i=f_{\rm low}(c_{\rm SALT, \it i})-f_{\rm high}(c_{\rm SALT, \it i});
	~~~
	m_{b,i}^{\rm corr}=m_{b,i}^{\rm public}+\Delta_i .
	\label{eq:massbias}
\end{equation}
Here $m_b$ is the standardized apparent magnitude and the index $i$ labels one Pantheon+ row.  This is an object-dependent update through the host-dependent model of \citet{Brout:2020msh} and the BEAMS with Bias Corrections methodology already encoded in the public Pantheon+ output, rather than the addition of one constant mass-step amplitude.  It does not rerun the full bias-correction simulations; it transfers each affected supernova from the public low-mass bias-correction branch to the corresponding high-mass branch.  Only the \texttt{m\_b\_corr} column is changed.

For every matching H0DN row we then apply
\begin{equation}
	m_{b,i}^{\rm new}=m_{b,i}^{\rm old}
	+\left(m_{b,i}^{\rm corr}-m_{b,i}^{\rm public}\right),
	\label{eq:patch}
\end{equation}
where ``old'' and ``new'' refer to the H0DN Hubble-flow magnitudes before and after the update.  The mean correction over all 114 reconstructed Pantheon+ rows is $0.07220$ mag.  The H0DN Hubble-flow file contains 52 affected rows, corresponding to 46 distinct supernova names, for which the mean correction is $0.06391$ mag.  We keep the H0DN covariance, redshifts, and peculiar-velocity treatment unchanged.  The low-redshift bias-correction uncertainties are smaller than $0.02$ mag and are nearly unchanged between the two host-mass branches, so no corresponding covariance update is applied.  This calculation therefore isolates the effect of the revised magnitudes and is the main result of this paper. None of the 55 H0DN calibrator supernovae is among the 114 corrected rows.  The calibrator rung is consequently unchanged in all the runs in this paper; no assumption about an unmeasured calibrator correction is required.

In a second calculation, we rebuild the full 277-row Hubble-flow rung from corrected Pantheon+ data.  We use every row carrying the public SH0ES Hubble-flow flag, the corrected magnitude vector, the corresponding public Pantheon+ statistical-plus-systematic covariance, and the public heliocentric and Hubble-diagram redshifts (not the H0DN network covariance and redshifts).  The heliocentric redshift, $z_{\rm HEL}$, is measured in the observer's frame.  The Hubble-diagram redshift, $z_{\rm HD}$, includes the corrections for the motion of the observer and the host galaxy that are used in the Pantheon+ Hubble diagram \citep{Carr:2021lcj,Peterson:2021hel}.  This second calculation uses the corrected Pantheon+ data consistently for the whole supernova rung, not only for the 52 changed magnitudes.

\subsection{DES-DOVEKIE and its magnitude offset}

The DES-DOVEKIE release provides a distance modulus, $\mu_{\rm DES}$, normalized with a nominal $H_0=70$ km s$^{-1}$ Mpc$^{-1}$, rather than the standardized apparent magnitude required by H0DN \citep{DES:2025sig}.  We select rows with $\mu_{\rm DES}$, $z_{\rm HEL}$, and $z_{\rm HD}$ within $0.0232\leq z_{\rm HD}\leq0.151$.  The 220 selected supernovae actually span $0.02509\leq z_{\rm HD}\leq0.15062$.  They comprise 197 external-survey and 23 DES-observed supernovae, of which 196 also occur in corrected Pantheon+; the fiducial comparison therefore mainly tests the DOVEKIE recalibration of a shared low-redshift sample.

\begin{figure*}[tb]
	\centering
	\includegraphics[width=\textwidth]{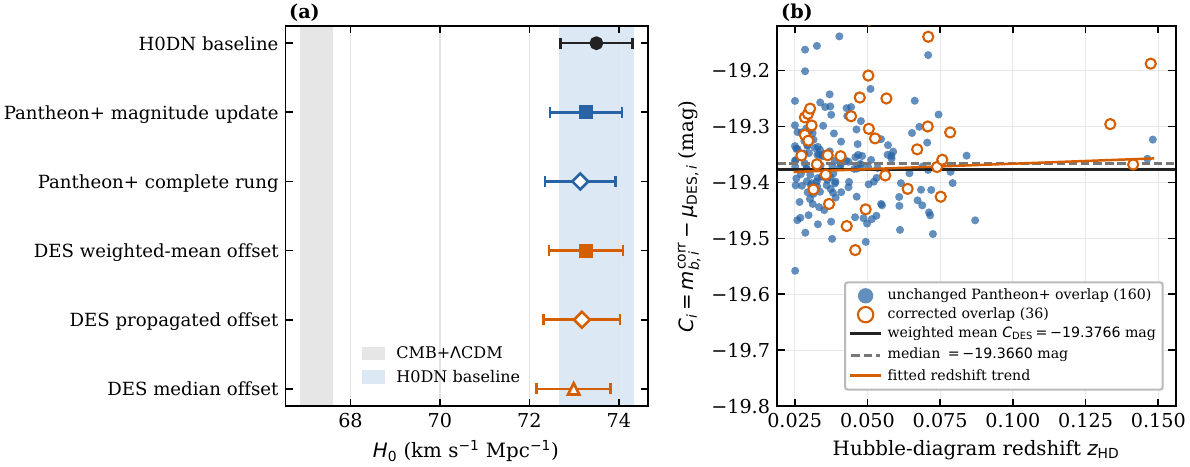}
	\caption{Hubble-constant results and the DES-DOVEKIE magnitude-offset check.  (a) Values of $H_0$ with $1\sigma$ uncertainties.  The first Pantheon+ case updates only the 52 affected magnitudes with the newly corrected Pantheon+ data, the second one replaces the complete 277-row H0DN Hubble-flow rung, its covariance, redshifts, and velocity corrections with the corrected Pantheon+ data.  The three DES cases use the fixed weighted-mean offset, its propagated covariance from Equation~(\ref{eq:vy}), or the fixed median.  The blue and gray bands show the H0DN baseline and $H_0^{\rm CMB,\Lambda CDM}=67.24\pm0.35$ km s$^{-1}$ Mpc$^{-1}$, respectively \citep{SPT-3G:2025bzu}.  (b) For the 196 supernovae shared by DES-DOVEKIE and corrected Pantheon+, $C_i=m_{b,i}^{\rm corr}-\mu_{{\rm DES},i}$.  Filled blue circles are the 160 unchanged overlaps; open orange circles are the 36 magnitude corrected overlaps.  The black, dashed gray, and orange lines show the weighted mean ($-19.3766$ mag), median ($-19.3660$ mag), and fitted redshift trend ($0.195\pm0.255$ mag per unit redshift); the scatter is 0.0823 mag.}
	\label{fig:summary}
\end{figure*}

An additive offset places these distance moduli on the H0DN magnitude convention.  Among the 196 overlaps, 36 have a magnitude changed by the host-property correction.  For each overlap we form
\begin{equation}
  C_i=m_{b,i}^{\rm corr}-\mu_{{\rm DES},i}, \qquad
  C_{\rm DES}=\frac{\sum_i w_i C_i}{\sum_i w_i},
  \label{eq:bridge}
\end{equation}
where $C_i$ is the offset for one supernova and $w_i=1/\sigma_{\mu,i}^2$ uses its released distance-modulus uncertainty, $\sigma_{\mu,i}$.  We find the weighted mean $C_{\rm DES}=-19.3766$ mag and construct
\begin{equation}
  m_{b,i}^{\rm DES,H0DN}=\mu_{{\rm DES},i}+C_{\rm DES}.
  \label{eq:desmb}
\end{equation}
The 196 individual offsets have an unweighted sample standard deviation or scatter of $0.0823$ mag.  Within this fiducial interval, their fitted redshift slope is $0.195\pm0.255$ mag per unit redshift, only $0.76\sigma$ from zero.  Figure~\ref{fig:summary}(b) shows this comparison.  As a simple check that the weighted mean is not selecting an unusually favorable zero point, we also use the ordinary median, $C_{\rm DES}^{\rm med}=-19.3660$ mag. $C_{\rm DES}$ absorbs both the fiducial absolute magnitude of the DES-DOVEKIE release and any mean difference in standardization between the two reductions --- the light-curve shape and color
coefficients, the host-mass step, the bias corrections, and the photometric calibration. These are not separable from the overlap data alone, which is why the offset is treated as a single
constant. Its adequacy rests on the absence of a resolved redshift trend rather than on the two reductions sharing a common standardization.

We note here that the stated unweighted sample standard deviation or scatter of $0.0823$ mag should not be confused with the offsets' full range of $0.5323$ mag, which is set by the magnitude difference of the two most extreme objects and does not describe the width of the overall distribution.  The scatter is relatively small because each offset compares two standardized measurements of the same supernova, so much of the intrinsic luminosity variation cancels; the remaining dispersion mainly reflects differences in calibration, light-curve modelling, and bias corrections between corrected Pantheon+ and DES-DOVEKIE.  The limited scatter therefore supports using a single additive offset to connect the two magnitude conventions. The measured scatter of $0.0823$ mag corresponds to a fractional luminosity distance dispersion of approximately $3.8\%$, using $\delta d_L/d_L\simeq(\ln 10/5)\delta\mu$.  For comparison, a scatter of $0.15$--$0.20$ mag, corresponding to roughly $7\%$--$9\%$ in distance, would be comparable to or larger than the dispersion of a standardized supernova sample and would raise concern that a single additive offset does not adequately connect the two reductions. Here, $d_L$ is the luminosity distance and $\mu \equiv 5\log_{10}(d_L/10\,{\rm pc})$ is the distance modulus.

For the H0DN redshift calculation, DES $z_{\rm HEL}$ is retained as the heliocentric redshift and DES $z_{\rm HD}$ is used as the effective flow-corrected redshift.  All additional H0DN velocity columns and the additional H0DN peculiar-velocity dispersion are set to zero, because the public DES $z_{\rm HD}$ values and statistical-plus-systematic covariance already include the adopted redshift and flow treatment; neither is therefore applied a second time. The released DES product is an inverse covariance.  We first invert the complete matrix, select the covariance rows and columns of the 220 supernovae, and then invert that selected covariance for H0DN.  

The first DES-DOVEKIE result treats $C_{\rm DES}$ as fixed.  In a second analysis, we also propagate the uncertainty in  $C_{\rm DES}$.  Let $\mathbf d$ be the full DES distance-modulus vector, $\mathbf r$ the corrected-Pantheon+ magnitudes of the overlaps, $F$ the matrix that selects the Hubble-flow rows, $S$ the matrix that selects the overlap rows, $\mathbf w$ the normalized weights in Equation~(\ref{eq:bridge}), and $\mathbf 1$ a column of ones.  The H0DN magnitude vector is then
\begin{equation}
 \mathbf y=F\mathbf d+\mathbf 1\,\mathbf w^{\mathsf T}(\mathbf r-S\mathbf d)
          =A\mathbf d+B\mathbf r,
 \label{eq:propagation}
\end{equation}
where $A=F-\mathbf 1\mathbf w^{\mathsf T}S$ and $B=\mathbf 1\mathbf w^{\mathsf T}$.  If $V_D$ and $V_R$ denote the public DES and Pantheon+ covariance matrices, respectively, the available covariance of $\mathbf y$ is
\begin{equation}
  V_y=A V_D A^{\mathsf T}+B V_R B^{\mathsf T}.
  \label{eq:vy}
\end{equation}
The superscript $\mathsf T$ denotes a vector or matrix transpose.  No joint DES--Pantheon+ cross-covariance has been made public, so Equation~(\ref{eq:vy}) sets that unprovided term to zero.  The two products overlap and should not be interpreted as statistically independent; this result propagates the covariance that is publicly available.  The resulting $V_y$ is symmetric and positive definite.

\section{Results} \label{sec:3}

\begin{table*}[tb]
	\centering
	\caption{H0DN results at the fiducial upper limit $z_{\rm HD}=0.151$.  $N_{\rm HF}$ is the number of Hubble-flow supernovae, $\Delta H_0$ is measured from the H0DN baseline, and $t_{\rm CMB}$ is the Gaussian difference from $H_0^{\rm CMB,\Lambda CDM}=67.24\pm0.35$ km s$^{-1}$ Mpc$^{-1}$.}
	\label{tab:main_results}
	\small
	\setlength{\tabcolsep}{7pt}
	\begin{tabular*}{\textwidth}{@{\extracolsep{\fill}}lccccc}
		\hline\hline
		Case & $N_{\rm HF}$ & $H_0$ & $\Delta H_0$ & $\chi^2/{\rm dof}$ & $t_{\rm CMB}$ \\
		& & (km s$^{-1}$ Mpc$^{-1}$) & (km s$^{-1}$ Mpc$^{-1}$) & & ($\sigma$) \\
		\hline
		H0DN baseline & 277 & $73.499\pm0.809$ & $+0.000$ & 0.988 & 7.10 \\
		\textbf{corrected Pantheon+: magnitude update only }& \textbf{277 }& \boldmath $73.264\pm0.806$ & \boldmath $-0.235$ &\textbf{ 0.990} &  \textbf{6.85 }\\
		corrected Pantheon+: complete Hubble-flow rung & 277 & $73.139\pm0.791$ & $-0.360$ & 0.999 & 6.82 \\
		DES-DOVEKIE: weighted-mean offset fixed & 220 & $73.267\pm0.828$ & $-0.232$ & 0.998 & 6.70 \\
		DES-DOVEKIE: offset covariance propagated & 220 & $73.175\pm0.854$ & $-0.323$ & 0.999 & 6.43 \\
		DES-DOVEKIE: median offset fixed & 220 & $72.991\pm0.825$ & $-0.508$ & 1.001 & 6.42 \\
		\hline
	\end{tabular*}
	\vspace{0.3em}
	\parbox{\textwidth}{\footnotesize \textit{Note.} All runs fix the deceleration parameter to $q_0=-0.55$ and the jerk parameter to $j_0=1$.  Here $\chi^2$ is the chi-square goodness-of-fit statistic and ``dof'' denotes degrees of freedom.  The reported $\chi^2/{\rm dof}$ is for the complete H0DN fit, which compresses the Hubble-flow sample to one intercept.  The offset is the constant that converts DES-DOVEKIE distance moduli to the H0DN apparent-magnitude convention.  Values of $\Delta H_0$ and $t_{\rm CMB}$ are calculated from the unrounded fit outputs.}
\end{table*}

\subsection{Main Results}

Figure~\ref{fig:summary}(a) and Table~\ref{tab:main_results} collect the H0DN fits.  The magnitude-only Pantheon+ update lowers the baseline by $0.235$ km s$^{-1}$ Mpc$^{-1}$, to

\begin{equation}\boldmath
	H_0 = 73.264\pm0.806 ~\mathrm{km}~\mathrm{s}^{-1} \mathrm{Mpc}^{-1},
\end{equation}

 which is our \textbf{main baseline result.} Rebuilding the complete corrected-Pantheon+ rung lowers it by $0.360$ km s$^{-1}$ Mpc$^{-1}$, to $73.139\pm0.791$ km s$^{-1}$ Mpc$^{-1}$.  These are two answers to two useful questions: the first isolates the revised host-property correction, while the second adopts all supernova information from the corrected Pantheon+ data.

The H0DN and Pantheon+ covariance matrices for these 277 rows are identical, and the post-update magnitude vectors agree within sub-millimagnitude file rounding.  The additional $0.125$ km s$^{-1}$ Mpc$^{-1}$ shift in the complete-rung result therefore comes almost entirely from replacing the H0DN redshifts and peculiar-velocity treatment with the Pantheon+ Hubble-diagram redshifts.

The mean correction of the 52 changed rows is not the quantity that sets the response of $H_0$.  The rows carry different covariance weights, and most H0DN rows do not change.  We therefore calculate the influence $g_i=\partial H_0/\partial m_{b,i}$ of each magnitude.  Summing $g_i\Delta_i$ predicts $-0.236$ km s$^{-1}$ Mpc$^{-1}$, in agreement with the exact H0DN change of $-0.235$ km s$^{-1}$ Mpc$^{-1}$.  The corresponding influence-weighted magnitude correction is only $0.0085$ mag, far below the coherent $0.236$ mag required to bring the full network to the adopted CMB central value.

The fixed weighted-mean DES offset gives $H_0=73.267\pm0.828$ km s$^{-1}$ Mpc$^{-1}$.  Propagating the available offset covariance gives $73.175\pm0.854$ km s$^{-1}$ Mpc$^{-1}$.  The latter uncertainty is larger, as expected, but the central value remains close to 73.  The median offset gives $72.991\pm0.825$ km s$^{-1}$ Mpc$^{-1}$, showing that the agreement is not a consequence of choosing the weighted mean. As a separate consistency check, restricting the DES color parameter to $|c_{\rm SALT}|<0.15$ leaves 198 supernovae in the Hubble-flow rung, and gives $73.338\pm0.836$ km s$^{-1}$ Mpc$^{-1}$ with $C_{\rm DES}$ fixed to the weighted mean.  Here $c_{\rm SALT}$ is the fitted supernova color used in the light-curve standardization. The $|c_{\rm SALT}|<0.15$ cut checks whether the 22 reddest objects, which may be more sensitive to dust, color standardization, and selection effects, materially affect $H_0$. They do not. All these checks change the result by less than 1$\sigma$ of the H0DN result.

To quote the early--late difference consistently, we define
\begin{equation}
 t_{\rm CMB}=\frac{H_0-H_0^{\rm CMB,\Lambda CDM}}
 {\sqrt{\sigma_{H_0}^2+\sigma_{H_0, \rm CMB}^2}},
 \label{eq:tension}
\end{equation}
where $H_0^{\rm CMB,\Lambda CDM}=67.24$ km s$^{-1}$ Mpc$^{-1}$, $\sigma_{H_0, \rm CMB}=0.35$ km s$^{-1}$ Mpc$^{-1}$, and $\sigma_{H_0}$ is the uncertainty of the H0DN results.  This CMB reference comes from a joint Planck, Atacama Cosmology Telescope, and South Pole Telescope fit within flat $\Lambda$CDM \citep{SPT-3G:2025bzu}, and is the reference adopted by H0DN \citep{H0DN:2025lyy}.  The baseline difference is $7.10\sigma$.  The two corrected-Pantheon+ results remain at $6.85\sigma$ and $6.82\sigma$; the DES-DOVEKIE results with fixed and uncertainty-propagated offset remain at $6.70\sigma$ and $6.43\sigma$ respectively.  Even the DES-DOVEKIE result with median-offset $H_0$ is $6.42\sigma$ above the CMB value.  The complete H0DN values of $\chi^2$ per degree of freedom lie between 0.988 and 1.001 for the cases in Table~\ref{tab:main_results}.

The reduction from 277 corrected-Pantheon+ Hubble-flow rows to 220 DES-DOVEKIE rows is visible in the uncertainty of the supernova intercept: $\sigma(\alpha_B)$ increases from $0.001811$ for the complete corrected-Pantheon+ rung to $0.002547$ for DES-DOVEKIE when $C_{\rm DES}$ is held fixed. The calibrator data, their covariance, and every non-supernova distance-network input are exactly unchanged; only the Hubble-flow contribution to the uncertainty budget is replaced. The final uncertainty therefore does not scale simply as $N_{\rm HF}^{-1/2}$, where $N_{\rm HF}$ is the number of Hubble-flow supernovae: it increases from $0.791$ to $0.828\,{\rm km\,s^{-1}\,Mpc^{-1}}$. The fixed-offset result is conditional on the weighted-mean value of $C_{\rm DES}$ and does not include the uncertainty associated with determining this transfer. Propagating the available DES-DOVEKIE and corrected-Pantheon+ covariances gives $\sigma(C_{\rm DES})=0.02309$ mag, with contributions of $0.01739$ mag from DES-DOVEKIE and $0.01520$ mag from corrected Pantheon+. Because $C_{\rm DES}$ is determined using many of the same DES-DOVEKIE measurements that form the Hubble-flow rung, its DES-side uncertainty is correlated with those measurements and cannot be added as an independent $0.02309$ mag zero-point error. Equation~(\ref{eq:vy}) propagates these correlations jointly, increasing $\sigma(\alpha_B)$ to $0.003040$ and the final $H_0$ uncertainty to $0.854\,{\rm km\,s^{-1}\,Mpc^{-1}}$.

\subsection{Extending the Hubble-flow samples}
\begin{figure*}[tb]
	\centering
	\includegraphics[width=0.86\textwidth]{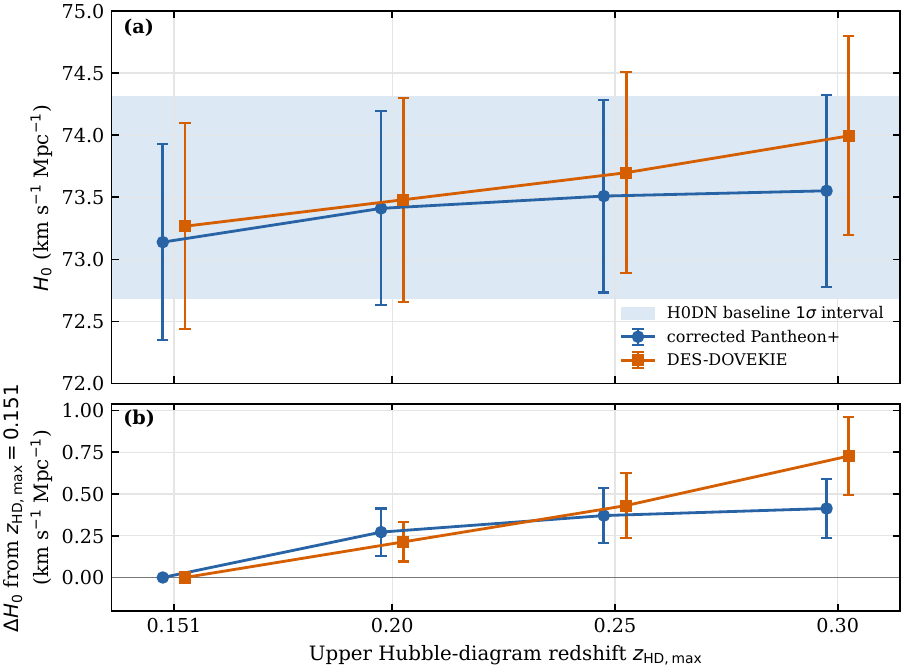}
	\caption{Dependence on the upper Hubble-diagram redshift, $z_{\rm HD,max}$.  The DES-DOVEKIE offset is measured once from the 196 overlaps at $z_{\rm HD,max}=0.151$ and is then held fixed.  (a) Values of $H_0$ for corrected Pantheon+ (full rung replacement case) and DES-DOVEKIE (fixed weighted-mean offset case); the pale band is the H0DN baseline $1\sigma$ interval.  (b) Changes relative to the first point.  Because the samples are nested, the lower-panel error bars use the covariance between redshift cuts rather than the individual $H_0$ errors added in quadrature.  Both sequences rise as more distant supernovae are included. }
	\label{fig:redshift}
\end{figure*}

The fiducial DES-DOVEKIE sample is strongly connected to the low-redshift reference set: 196 of its 220 supernovae enter the offset measurement.  We therefore repeat the calculation at upper Hubble-diagram redshifts of  $z_{\rm HD, max}=0.20$, 0.25, and 0.30.  The same 196-object offset is held fixed, so the added supernovae test the relative distance scale rather than refitting its zero point. The H0DN luminosity-distance relation used here is third order in
redshift, its quadratic cosmographic correction multiplying the leading factor of $z$. The adopted values $(q_0, j_0) = (-0.55, 1)$ are exactly those of a spatially flat $\Lambda$CDM model with present matter density parameter $\Omega_{\rm m} = 0.30$, for which $j_0 = 1$ and $q_0 = \tfrac{1}{2}\Omega_{\rm m} - \Omega_\Lambda = -0.55$, where $\Omega_\Lambda = 1 - \Omega_{\rm m}$. Over $0.151 \leq z_{\rm HD} \leq 0.30$ the truncated relation departs from that model by only about $1$ millimagnitude in relative distance modulus. The upper cut also remains well inside the $z \simeq 1$ convergence radius of the standard redshift expansion \citep{Cattoen:2007sk}. The redshift extension is therefore well motivated as a consistency check.

For corrected Pantheon+ (we are considering the full Hubble-flow rung replacement case including magnitudes, covariance, redshifts, velocities), the first row retains the 277 objects carrying the SH0ES Hubble-flow flag; the higher cuts add noncalibrator supernovae that pass the public light-curve cuts.  This gives 659 corrected-Pantheon+ and 403 DES-DOVEKIE supernovae by $z_{\rm HD}=0.30$; 206 of the latter were observed by DES itself. The resulting dependence of $H_0$ on the upper redshift limit, together with the changes relative to the fiducial cut, is shown in Figure~\ref{fig:redshift}.

The two samples move the $H_0$ in the same direction.  Between the redshift upper cuts of 0.151 and 0.30, the $H_0$ from corrected Pantheon+ rises by $0.413\pm0.179$ km s$^{-1}$ Mpc$^{-1}$  and DES-DOVEKIE rises by $0.728\pm0.233$ km s$^{-1}$ Mpc$^{-1}$.  The uncertainties come from 5000 joint Gaussian draws with the covariance between the nested samples retained; a separate uncertainty is calculated at every cut.  The rises differ by $0.314\pm0.294$ km s$^{-1}$ Mpc$^{-1}$, only $1.07\sigma$ from zero; their unavailable cross-covariance is omitted.  Their agreement does not point to an effect unique to DES-DOVEKIE, but it does not distinguish expansion-history sensitivity from calibration or selection effects common to the overlapping samples. Noticeably, neither sequence moves toward the CMB value. The central values of the rises are smaller than the $1\sigma$ error of the H0DN baseline result, and thus do not necessitate invocation of new physics. The overlap between the samples prevents us from distinguishing expansion-history sensitivity from calibration or selection effects. Crucially, both sequences move away from, rather than toward, the CMB+$\Lambda$CDM value and therefore do not alleviate the Hubble tension. The redshift interval probed here is also where an evolving dark-energy equation of state would first enter the Hubble-flow rung. The supernova revisions that underlie the current cross-sample evidence for dynamical dark energy therefore leave the local distance scale essentially unchanged.

The intermediate points follow the same progression.  At upper cuts of 0.20 and 0.25, corrected Pantheon+ gives $73.411\pm0.781$ and $73.510\pm0.777$ km s$^{-1}$ Mpc$^{-1}$ from 400 and 539 supernovae, while DES-DOVEKIE gives $73.480\pm0.822$ and $73.698\pm0.809$ km s$^{-1}$ Mpc$^{-1}$ from 245 and 314 supernovae.  At 0.30 the corresponding values are $73.552\pm0.774$ for the corrected Pantheon+ and $73.995\pm0.801$ km s$^{-1}$ Mpc$^{-1}$ for DES-DOVEKIE from 659 and 403 supernovae respectively.  The correlated uncertainties of the changes at 0.20 and 0.25 are, respectively, 0.141 and 0.166 km s$^{-1}$ Mpc$^{-1}$ for corrected Pantheon+ and 0.117 and 0.196 km s$^{-1}$ Mpc$^{-1}$ for DES-DOVEKIE.

As a final consistency check, we retain only the corrected Pantheon+ magnitude shifts while leaving the original H0 Distance Network (H0DN) covariance, redshifts, and 2M++ peculiar-velocity treatment unchanged, and raise the lower Hubble-diagram redshift limit from $z_{\rm HD}=0.0232$ to $0.05$, keeping $z_{\rm HD}\leq0.151$. For the $240\,{\rm km\,s^{-1}}$ velocity dispersion adopted by H0DN, the approximate contribution to the distance modulus follows from the standard low-redshift propagation of peculiar velocities into magnitude uncertainties \citep{Hui:2005nm,Davis:2010jq}:
\begin{equation}
	\sigma_{\mu,v}\simeq\frac{5}{\ln 10}\frac{\sigma_v}{cz}.
\end{equation}
Here, $\sigma_{\mu,v}$ is the uncertainty in the distance modulus, $\mu$, induced by the line-of-sight peculiar velocity; the quantity $\sigma_v$ is the adopted one-dimensional peculiar-velocity dispersion, and $c$ is the speed of light. From the above equation it follows that $\sigma_{\mu,v}$ decreases from $0.075$ mag, or $3.4\%$ in distance, at $z=0.0232$ to $0.035$ mag, or $1.6\%$, at $z=0.05$. Peculiar velocities are therefore subdominant to the typical $\sim0.1$ mag scatter of standardized Type Ia supernovae above the higher threshold.  The $z_{\rm HD}> 0.05$ cut retains 91 of the 277 Hubble-flow rows and gives $H_0=72.896\pm0.858\,{\rm km\,s^{-1}\,Mpc^{-1}}$, only $0.367\,{\rm km\,s^{-1}\,Mpc^{-1}}$ below the fiducial corrected-Pantheon+ result and still $6.11\sigma$ above $H_0^{\mathrm{CMB},\Lambda\mathrm{CDM}}=67.24\pm0.35\,{\rm km\,s^{-1}\,Mpc^{-1}}$. Thus, the high value of $H_0$ is retained after removing the supernovae most susceptible to peculiar-velocity uncertainties.

\section{Conclusion}\label{sec:4}

Corrected Pantheon+ and DES-DOVEKIE alter the $H_0$ Distance Network (H0DN) supernova rung in different ways, yet they lead to the same broad result.  Updating only the corrected Pantheon+ magnitudes gives $H_0=73.264\pm0.806$ km s$^{-1}$ Mpc$^{-1}$. This is the most up-to-date Hubble constant value for the astrophysics community and the main baseline result of our paper. The Hubble tension remains significant at 6.85$\sigma$ from the CMB+$\Lambda$CDM determination of $H_0^{\rm CMB,\Lambda CDM}=67.24\pm0.35$ km s$^{-1}$ Mpc$^{-1}$.

Rebuilding the complete Hubble-flow rung using the public Pantheon+ covariance, redshifts, and velocity corrections gives $73.139\pm0.791$ km s$^{-1}$ Mpc$^{-1}$.  Replacing the Hubble-flow rung with DES-DOVEKIE supernovae gives $73.267\pm0.828$ km s$^{-1}$ Mpc$^{-1}$ when the overlap offset is fixed and $73.175\pm0.854$ km s$^{-1}$ Mpc$^{-1}$ when the offset's available uncertainty is propagated.  Using the median-offset results in  $H_0 = 72.991\pm0.825$ km s$^{-1}$ Mpc$^{-1}$, which is the lowest central value among the cases shown, but the Hubble tension still remains $6.42\sigma$. As per our knowledge, this is the first use of DES-Dovekie in a local distance-ladder determination of $H_0$.

The correction to the Pantheon+ data does not change the magnitudes of any calibrator supernova, but 52 of the 277 Hubble-flow rows do change. However, their influence-weighted correction is only $0.0085$ mag.  The full H0DN network would require a coherent $0.236$ mag change to reach the CMB central value.  Extending either corrected Pantheon+ or DES-DOVEKIE to redshift 0.3 raises rather than lowers $H_0$, with no significant difference between the two rises.  Within the public H0DN framework and the fixed $(q_0,j_0)=(-0.55,1)$ convention, these supernova updates therefore leave the Hubble constant close to 73 km s$^{-1}$ Mpc$^{-1}$ and do not resolve the early--late disagreement.

These results acquire broader significance in light of the recent convergence among the three principal supernova samples. When combined with DESI BAO and CMB measurements, corrected Pantheon+, Union3.1, and DES-DOVEKIE yield substantially more consistent constraints, with a $3.2$--$3.4\sigma$ preference for dynamical dark energy \citep{Hoyt:2026fve}. We find that the corresponding supernova updates have only a minor effect on the local distance scale: both corrected Pantheon+ and DES-DOVEKIE leave the H0DN determination close to $H_0 \simeq 73\,\mathrm{km\,s^{-1}\,Mpc^{-1}}$. Thus, revisions that are important for the current evidence for dynamical dark energy do not materially reduce the Hubble tension.

\begin{acknowledgments}
The author thanks Taylor Hoyt and David Rubin for kindly sharing the corrected Pantheon+ data file. The author also acknowledges support from grant Nos. \\ AS-IA-112-M04, NSTC   112-2112-M-001-027-MY3, and I-IAA-ROY.  
\end{acknowledgments}

\bibliographystyle{aasjournal}
\bibliography{references_v3}

\end{document}